# WEBPAGE LOAD SPEED: ASP.NET VS. PHP

By
**TIMUR MIRZOEV**

*Georgia Southern University; Lawton Sack, Georgia Southern University.*

**ABSTRACT**

As data transmission speeds over the Internet continue to increase, it is necessary to research and identify technologies that can take advantage of the increased speeds by enhancing the loading speed of webpages. One area of consideration is found in the type of framework that is utilized for a website. There are numerous frameworks that can be chosen from to be used to support a website, each with their distinctive advantages. There are many differing opinions that have been proffered on which framework should be used to fully realize the optimization of page load speed. This manuscript examines and implements testing methods for two popular frameworks, Active Server Pages .net and PHP, to make a final determination of which framework is most beneficial for webpage load speeds.

INTRODUCTION

A web developer must carefully choose which framework will be utilized to support the website they are creating, as it will have an impact on all facets of the design, installation, operation, and maintenance of the site. A 2010 study of approximately 6.7 million domains utilizing a determinable framework revealed that 59 percent of the domains are used to PHP while 34 percent used to ASP.net [1]. In stark comparison, the third place finisher amongst the domains was Perl, which accounted for only 4 percent of the identified domains [1]. These numbers strongly indicate that the two major choices for web developers are PHP and ASP. net, though they not able to provide a solid reasoning behind why one framework was chosen by a developer over another framework.

While PHP and ASP. net have offered numerous, yet distinct advantages that played a part in the decision making of web developers in which framework should be utilized, the advantages have not directly addressed the impact that the choice between the two ultimately has upon the load speeds for web pages on a web site. Unfortunately, many people have fallen into the temptation of using personal bias to make a proclamation that a particular framework is faster than another. A clear proclamation of dominance by one framework, if one exists, It should be made after research and testing have been performed. This paper takes an examination of both frameworks to determine if the numerous proclamations that ASP. net is faster than PHP are indeed accurate.

1. The Importance of Speed

In 22, 2012, there were approximately 6.89 billion pages indexed on the World Wide Web [2]. Each of these billions of pages requires the support of a web framework to handle the presentation and responsiveness of the web page by processing code and reacting to the responses and requests generated by the user, a browser, servers, and other devices. A framework must be robust without sacrificing performance and negatively impacting the speed of loading a web page. Unfortunately, there is not a perfect web framework available, as different web frameworks put a focus on different problems while providing unique solutions to those problems [3]. Though the design and support are fairly different between PHP and ASP. net, both frameworks attempt to be a reliable solution for web designers.

PHP is one of the most widely used programming languages in the world [4] that is run on an open-source framework that is capable of running on each of the major server platforms, including Unix, Linux, Windows, and Macintosh systems [5]. This versatility even has been noted by Microsoft, the developers of ASP. net, in an acknowledgement that PHP has become the de facto alternative to ASP. net for website development due to the fact that ASP. net is not officially supported on Linux and Unix systems, which run a large majority of the web's servers [6]. The versatility in regards to the server system is not the only reason that many web developers choose PHP, though, as PHP is also extremely popular due its simplicity of use, its stability, its widespread support from the open-source community, and its overall flexibility through the use of extensions, server interfaces, database interfaces, and other available modules [5]. Popular web sites that utilize PHP include Facebook, Google, Yahoo!, Wikipedia, and WordPress [7].

ASP.net is a free, server-side technology that is owned and developed by Microsoft that only officially runs only on Windows servers and part of Microsoft's .NET framework [8]. But some feel that it is a limitation for ASP.net to not be an open-source framework, others feel it is a strong advantage to have a multi-billion dollar company overseeing the development and maintenance of the ASP.net framework. Microsoft has included in ASP.net the ability for the web programmer to choose from both script languages and .NET languages, such as C#, Visual Basic, and J#, within the ASP.net framework [9]. This is a powerful advantage that has led to its strong popularity amongst web developers. Some additional advantages to ASP.net include a reduction in the amount of code necessary to build large and secure applications, a rich tool box and designer in the Visual Studio development environment, and the overall ease of deployment on a Windows server [9]. The two major sites that utilize ASP.net include MSN.com and Live.com, which are both owned by Microsoft [7].

Obviously, each framework has disadvantages that have dissuaded by some programmers from choosing one or both of the frameworks. In the case of PHP, it has been argued that its framework is not suitable for large projects and complex sites [4] and it often takes additional languages, such as JavaScript, Perl, or Java, to make a PHP site more useable [10]. For ASP.net, the higher hardware, software, and time costs have been the most repeated disadvantages when compared to the free and less time-intensive PHP [11].

Internet users are typically more concerned with the speed that a web page loads and not the framework that actually supports the site they are visiting. In fact, Harry Shum, a computer scientist and speed specialist at Microsoft, states that "two hundred fifty milliseconds, either slower or faster, is close to the magic number now for competitive advantage on the Web [12]." A web site that loads a page at a speed of one quarter of second slower than a competitor will be a distinct disadvantage and lose visitors to the faster competitor [12]. Further, a Forrester Research study showed that as many as 40 percent of consumers would actually abandon a page if took more than three seconds to load the page [13]. This drives web designers and programmers search for a web framework that can provide a speed advantage over its competitors without having a sacrifice reliability.

Though it has been used less widely than PHP, there have been numerous proclamations made in the past that the ASP.net framework was universally faster than PHP [14] and that compiled code, such as ASP.net, ran faster than interpreted languages like PHP [15]. The overall popularity of PHP has caused reasonable doubt to arise on the accuracy of these claims, though. Are the programmers that choose to use PHP simply distracted by their own personal bias or is it possible that the purported claims of ASP.net's superiority in web page load speeds is actually non-existent? An unbiased experimentation must take place to settle, at least for now, if there is indeed a speed advantage present in ASP.net.

2. The Race

Due to the persistence of speculation between whether ASP.net or PHP was the faster framework to implement, it was necessary to implement testing methodologies to attempt to make an unbiased determination if either framework provided a clear advantage in webpage load speed. Research indicated that numerous testing apparatus and methodologies were available to produce accurate and reliable test results. The final determination was made to test the two frameworks on a single local machine in order to avoid common Internet issues such as inconsistent bandwidth, packet loss, and dissimilar routing paths that could have caused the testing results to be skewed.

3. Hardware Setup

All testing procedures were performed on a computer system that utilized an Intel Core i7-2600K CPU processor running at 3.40 GHz with 16.0 GB of RAM. The operating system of the computer was Windows 7 Professional 64-Bit with Service Pack 1 installed. Each of the ASP.net pages was run locally on Internet Information Service (IIS) 7.5.7600.16385 using ASP.NET 2.0.50727.5456. Both PHP pages were run locally on Wamp Server using Apache 2.2.2 and PHP 5.3.13. Load times of the web pages were measured using the lori (Life-of-request info) 0.2.0.20080521 extension within Firefox 13.0.1.

The lori Firefox extension was designed by Hung Le as a tool to be used by web developers to determine the length of time taken to completely load a webpage within the Firefox browser [16]. The extension was easily installed as a browser add-on in Firefox and it immediately available for use upon restarting the browser. A status bar was added by the lori extension in the lower right-hand corner of the Firefox browser that displayed various and useful statistics when a page was fully loaded, which included:

1. Time to First Byte (TTFB), which measured the time for the browser to receive the first byte from the server;

2. Time To Complete (TTC), which measured the entire time taken to completely load and draw the web page;

3. Page size, which included the size of information received from both the server and the cache; and

4. Number of requests made to fetch the content of the page (lori, n.d.).

For the testing purposes during this experimentation, the Time To Complete (TTC) was the statistic utilized, as the other statistics were irrelevant for this process.

4. Testing Cases

In order to test the load times for both PHP and ASP.net, the following web page files were created:

5.3KB ASP.net web page file named about.aspx that called the following external files:

- 3 JavaScript files
- 1 style sheet
- 1 CSS image
- 3 additional images
- 1 favicon

5.3KB PHP web page file named about.php that called the following external files:

- 3 JavaScript files
- 1 style sheet
- 1 CSS image
- 3 additional images
- 1 favicon

A 512B file named test.aspx that read a 166KB text file named Persons.txt that contained 10,000 lines of alphanumeric sequences that each had 15 characters. Each sequence was displayed in the browser with a line break between each sequence.

A 291B file named test.php that read a 166KB text file named Persons.txt that contained 10,000 lines of alphanumeric sequences that each had 15 characters. Each sequence was displayed in the browser with a line break between each sequence.

5. The Procedure

Two different tests that shared a similar procedure comprised the experimentation to test the web page load speeds of each framework. The first testing session consisted of manually entering the file path of a web page file (either about.aspx or about.php) into the address bar of the Fire Fox browser with the lori extension running. Once the page fully loaded, the total loading time reported by the lori extension was recorded into an Excel spreadsheet. After each page was loaded and the time recorded, the browser's cache was forcefully cleared and the page reloaded using the CTRL + F5 key combination. Both the ASP.net file (about.aspx) and the PHP file (about.php) were loaded for 25 iterations each. The second testing procedure was identical to the first, except that the test.aspx file was used for ASP.net and the test.php file was used for PHP and the test was performed two days later. In all, fifty iterations were run each on the ASP.net and PHP frameworks.

## 6. The Race Results

The purpose of the two tests was to form a determination if ASP.net actually provided a clear and distinctive speed advantage over PHP for loading webpages. It was necessary to provide a set of procedures that would accurately and consistently produce the results throughout both the tests were unbiased towards either framework. These objectives were met by performing the tests on a local machine using the IIS server for ASP.net pages and the WAMP server for PHP pages, both which are designed for serving the respective type of page. A highly recommended and long-standing Fire fox extension, lori, was chosen and utilized to record the time to fully load and display the web pages on each framework.

The analysis of the results obtained during the two testing sessions revealed that neither ASP.net nor PHP had a clear and distinctive speed advantage in terms of page load times, despite the expectation that ASP.net would load much faster as indicated in various reports. Though the two PHP files would score the fastest load times for both sessions at 0.181 seconds in session 1 and 1.281 seconds in session 2 (see Table 1), PHP also had one of the two slowest loading times at 0.304 seconds in session 1.

While the ASP.net page (about.aspx) secured the lower mean time than the PHP page (about.php) during the first testing session, it did so by just 11.2 milliseconds (see Table 1). This minor victory was short-lived, though, as the second testing session (see Table 1) showed that the PHP file (test.php) loaded an average of 13.56 milliseconds faster than the ASP.net page (test.aspx). Both of these results are well below the 250 millisecond threshold set by Harry Shum and thus would not indicate a competitive advantage for either framework[12].

Final Race Analysis

Whenever one technology boasts of providing a particular advantage over another competing technology, as many proponents of ASP.net made in terms of ASP.net being faster than PHP, it is necessary that data is present to unequivocally support the boasting. Also, it was crucial that the results from the experimentation were produced in a fashion that was both unbiased and easily reproduced by future testers and researchers. This manuscript's testing procedures were carefully designed to create a speed competition between the two frameworks that took place in a controlled environment that allowed for each framework to load the respective web pages in a fair test.

A lot of hype and hyperbole was found to be present in the research performed before the testing sessions took place. Unfortunately, there were many proclamations of one framework being faster than another, there was very little testing performed to actually back up the proclamations. The results of the two testing sessions performed for this manuscript cannot be ignored based upon prior proclamations of ASP.net supremacy in page load speeds, as they were produced, recorded, and calculated in a fair and scientific manner. The numbers obtained as a result of the two testing sessions revealed that the reported page load speed advantage of ASP.net over PHP was actually non-existent. Further, it was shown that neither ASP.net nor PHP showed any type of advantage in overall consistency of speed during the trials. Therefore, the consistency of speed and the actual load times should be ultimately irrelevant in the determination of which framework should be chosen by a web developer, as the myth of the supremacy of ASP.net has been disproved.

|  | Testing Session 1 | | Testing Session 2 | |
|---|---|---|---|---|
| Iteration | about.aspx | about.php | test.aspx | test.php |
| 1 | 0.222 | 0.202 | 1.438 | 1.317 |
| 2 | 0.198 | 0.209 | 1.339 | 1.337 |
| 3 | 0.294 | 0.190 | 1.304 | 1.337 |
| 4 | 0.214 | 0.226 | 1.323 | 1.296 |
| 5 | 0.193 | 0.198 | 1.294 | 1.301 |
| 6 | 0.243 | 0.181 | 1.370 | 1.317 |
| 7 | 0.209 | 0.208 | 1.305 | 1.313 |
| 8 | 0.197 | 0.231 | 1.288 | 1.332 |
| 9 | 0.228 | 0.229 | 1.312 | 1.381 |
| 10 | 0.195 | 0.287 | 1.305 | 1.310 |
| 11 | 0.189 | 0.295 | 1.304 | 1.281 |
| 12 | 0.189 | 0.267 | 1.318 | 1.370 |
| 13 | 0.229 | 0.225 | 1.328 | 1.328 |
| 14 | 0.249 | 0.304 | 1.313 | 1.314 |
| 15 | 0.200 | 0.203 | 1.371 | 1.302 |
| 16 | 0.191 | 0.214 | 1.299 | 1.314 |
| 17 | 0.197 | 0.209 | 1.405 | 1.298 |
| 18 | 0.245 | 0.197 | 1.388 | 1.291 |
| 19 | 0.202 | 0.270 | 1.404 | 1.335 |
| 20 | 0.205 | 0.237 | 1.329 | 1.331 |
| 21 | 0.245 | 0.257 | 1.385 | 1.337 |
| 22 | 0.224 | 0.207 | 1.379 | 1.300 |
| 23 | 0.223 | 0.223 | 1.319 | 1.427 |
| 24 | 0.207 | 0.229 | 1.386 | 1.332 |
| 25 | 0.239 | 0.209 | 1.300 | 1.366 |
| Mean | 0.21708 | 0.22828 | 1.34024 | 1.32668 |
| Standard Deviation | 0.0252915 | 0.0335006 | 0.042762 | 0.032263 |
| Min | 0.189 | 0.181 | 1.288 | 1.281 |
| Max | 0.294 | 0.304 | 1.438 | 1.427 |
| Max & Min Difference | 0.105 | 0.123 | 0.15 | 0.146 |

Table 1. Test Results in Seconds

According to the readings performed before the experimentation took place, there was an expectation of a significant difference in the speed calculations that favored the ASP.net framework. If the ASP.net framework was indeed faster and more consistent, then the testing sessions should have produced lower mean and median load times, closer minimum and maximum times than those for PHP, and lower standard deviations for ASP.net than the PHP framework. The results produced during the testing sessions and the calculations performed on that resulting data showed a much different picture than was expected, though. The test results clearly showed that neither framework could show an advantage in the mean and median speed loads times, the length of time span between the maximum and minimum load times, and in the standard deviations.

When the mean and median times are calculated and examined for the test result data, it is clear that neither offers any type of significant speed advantage. The mean times for ASP.net were .21708 seconds for session 1 and 1.34024 seconds for session 2, while PHP produced mean times of .22828 seconds and 1.32668 seconds in the two sessions. Since the median is a better indicator of central tendency than the mean, it is important to also look at the median time when dealing with such small time measurements [17]. ASP.net produced a median time of .209 seconds for session 1 and 1.323 seconds for session 2, compared to the median time of .223 seconds and 1.317 seconds for the PHP framework in sessions 1 and 2, respectively. Both the mean and median times shows that the both frameworks could possibly show a very slight advantage in a particular instances, but that neither could consistently produce an advantage in all circumstances or even at a level that could be noticed by users.

The results of the two testing sessions also demonstrated the fact that both frameworks produced extremely similar results in the minimum and maximum load times during the testing sessions. In test session 1, only eight thousandths of second separated the two frameworks at their fastest and only one hundredth of a second separation between the two at their slowest load time. In similar fashion, the test session 2 results showed a separation of seven thousandths of a second during the lowest load times and 1.1 hundredths of a second during the slower load times. Further, when the difference between the maximum and minimum load times for each framework were taken into consideration, the largest difference was only 1.8 hundredths of second, which occurred in favor of ASP.net during testing session 1. These results clearly prove that a clear and distinctive speed advantage was non-existent even when the extremes were taken into consideration.

The extremely low standard deviations calculated on the data collected during the testing sessions showed that each of the times recorded was very close to mean of that data. This demonstrated a high-level of consistency and reliability in the load times for both frameworks. It could be argued, though, that with standard deviations of 0.0335006 and 0.032263 produced on the PHP framework were even more consistent in its page load speeds when taken in comparison to the standard deviations of 0.0252915 and 0.042762 for ASP.net. However, the results for testing session 1 were closer together for ASP.net than PHP, so a clear distinction cannot made for a more consistent load time for either framework.

## Conclusions and Recommendations

Research and testing has proven that the load speed of web pages is a critical factor in the effort to retain visitors to a particular website. While webmasters have been offered many framework choices throughout the years, the two most popular frameworks were PHP and ASP.net at the time of this manuscript's experimentation. It will continue to be a necessity for designers, engineers, and researchers to routinely test the boundaries of these framework technologies to determine which one, if either, is the fastest framework to host a webpage on. By building a fair and equitable testing scheme, the experimentation that was performed for this manuscript made an important step forward in adding to the results that can aid these researchers and developers as they perform different tests on various computers and servers.

The design and implementation of unbiased testing routines was necessary to pit the ASP.net and PHP frameworks against each other in order to determine, if the ASP.net framework provided a speed advantage over PHP. A race consisting of two rounds of tests took place with the Firefox browser serving as the track, a WAMP server (PHP) and an IIS server (ASP.net) as the race car engines, the web pages acting as the race car body itself, and the lori extension acting as the timing mechanism. Once the dust settled from this particular set of two races, the data results clearly indicated that neither PHP nor ASP.net could be conclusively proven to have any type of speed advantage. Future research and testing will be important and necessary to verify the results of this manuscript, though.

The experimentation for this project was limited to one specific Windows 7 computer running one particular version of the WAMP server for PHP pages and one version of IIS for ASP.net web pages. Further, the tests only utilized two sets of pages during a limited number of iterations for each framework. This limited scope provided important and useful data, but it was only a small piece of the future testing that needs to occur in the future. In order to build upon the results produced by this experimentation, researchers will find it to be both necessary and beneficial that different types of computers and servers with different processors speeds and RAM configurations are used under varying process loads that could impact the access, retrieval, and displaying of web pages on both local and remote servers. Further, it would be important that different browsers and timing mechanisms are used on various operating systems to fully gauge that the results of this test were not isolated anomalies and are truly reflective of webpage load speeds in general. It is also important that other frameworks be considered for future testing and research. While ASP.net and PHP have proven to be extremely popular frameworks, it is very possible that other technologies may exist that actually provide a speed advantage over both ASP.net and PHP.

## ABOUT THE AUTHORS

TIMUR MIRZOEV is a professor of Information Technology Department at Georgia Southern University, College of Engineering and Information Technology. Dr. Mirzoev received his Ph.D. in Technology Management from Indiana State University in 2007. Dr. Mirzoev holds an MBA, an MS and two Bachelor's degrees. Dr. Mirzoev heads the International VMware IT Academy Center and EMC Academic Alliance at Georgia Southern University. His research interests include cloud computing, virtualization, storage areas in which he is commercially certified.

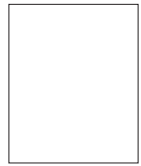

LAWTON SACK is an undergraduate student at Georgia Southern University in Statesboro, GA. He is pursuing a B.S. degree in Information Technology. His interests are in computer programming, digital publication, reading, and watching sports.

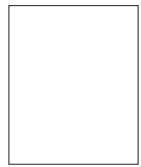